\documentclass[english,prl,twocolumn,showpacs,preprintnumbers,amsmath,amssymb,floatfix]{revtex4}

\usepackage{graphics,graphicx,color,hyperref}

\begin{document}

\title{Quantum Monte Carlo Calculation of the Binding Energy of Bilayer
  Graphene}

\author{E.\ Mostaani}

\affiliation{Department of Physics, Lancaster University, Lancaster LA1 4YB,
  United Kingdom}

\author{N.\ D.\ Drummond}

\affiliation{Department of Physics, Lancaster University, Lancaster LA1 4YB,
  United Kingdom}

\author{V.\ I.\ Fal'ko}

\affiliation{Department of Physics, Lancaster University, Lancaster LA1 4YB,
  United Kingdom}

\date{\today}

\begin{abstract}
We report diffusion quantum Monte Carlo calculations of the interlayer binding
energy of bilayer graphene. We find the binding energies of the AA- and
AB-stacked structures at the equilibrium separation to be $11.5(9)$ and
$17.7(9)$ meV/atom, respectively. The out-of-plane zone-center optical phonon
frequency predicted by our binding-energy curve is consistent with available
experimental results.  As well as assisting the modeling of interactions
between graphene layers, our results will facilitate the development of van
der Waals exchange--correlation functionals for density functional theory
calculations. 
\end{abstract}

\pacs{61.48.Gh, 71.15.Nc, 02.70.Ss}

\maketitle

van der Waals (vdW) interactions play a crucial role in a wide range of
physical and biological phenomena, from the binding of rare-gas solids to the
folding of proteins. Significant efforts are therefore being made to develop
computational methods that predict vdW contributions to energies of adhesion,
particularly for materials such as multilayer graphene. This task has proved
to be challenging, however, because vdW interactions are caused by nonlocal
electron correlation effects.  Standard first-principles approaches such as
density functional theory (DFT) with local exchange--correlation functionals do
not describe vdW interactions accurately. One technique for including vdW
interactions in a first-principles framework is to add energies obtained using
pairwise interatomic potentials to DFT total energies; this is the so-called
DFT-D scheme \cite{Grimme_2004,Hasegawa_2007,Grimme_2010,Tkatchenko_2009}. The
development of vdW density functionals (vdW-DFs) that can describe vdW
interactions in a seamless fashion is another promising approach
\cite{Rydberg_2003,Dion_2004,Chakarova_2006,Thonhauser_2007}.  DFT-based
random-phase approximation (RPA) calculations of the correlation energy
\cite{Lebegue_2010,Olsen_2013} provide a more sophisticated method for
treating vdW interactions; however, RPA atomization energies are typically
overestimated by up to 15\% for solids \cite{Eshuis_2012,Ren_2012}, and hence
the accuracy of this approach is unclear.  Symmetry-adapted perturbation
theory based on DFT allows one to calculate the vdW interactions between
molecules and hence, by extrapolation, between nanostructures
\cite{sapt}. Finally, empirical interatomic potentials with $r^{-6}$ tails may
be used to calculate binding energies \cite{Girifalco_1956,Girifalco_2000},
although such potentials give a qualitatively incorrect description of the
interaction of metallic or $\pi$-bonded two-dimensional (2D) materials at
large separation \cite{Dobson_2006}.

A key test system for methods purporting to describe vdW interactions between
low-dimensional materials is bilayer graphene (BLG)\@.  Several theoretical
studies have used methods based on DFT to calculate the binding energy (BE) of
BLG\@.  Some of the results are summarized in Table \ref{table:compare_BE},
but there is very little consensus. In this work we provide diffusion quantum
Monte Carlo (DMC) data for the BE of BLG and the atomization energy of
monolayer graphene (MLG), which we have extrapolated to the thermodynamic
limit. We find the DMC BE of BLG to be somewhat less than the BEs predicted by
DFT-D, although the latter vary significantly from scheme to scheme.  The DMC
method is the most accurate first-principles technique available for studying
condensed matter. Our data can therefore be used as a benchmark for the
development of vdW functionals.

\begin{table}
\caption{BE of BLG (both AA- and AB-stacked) obtained in recent theoretical
  studies. The layer separations $d$ quoted in the table are the ones used in
  the calculations, not necessarily the optimized bond length for the given
  method.  ``SAPT(DFT)'' and ``DFT-LCAO-OO'' denote symmetry-adapted
  perturbation theory based on DFT and linear combination of atomic
  orbitals-orbital occupancy based on DFT, respectively.
\label{table:compare_BE}}
\begin{tabular}{lcr@{}lr@{}l} \hline \hline

Stacking & Method & \multicolumn{2}{c}{$d$ ({\AA})} & \multicolumn{2}{c}{BE
  (meV/atom)} \\
\hline
AA  & vdW-DF \cite{Lebedeva_2011}   & $3$&$.35$    & $10$&$.4$    \\

AA  & DFT-D \cite{Lebedeva_2011}    & $3$&$.25$    & $31$&$.1$ \\

AA & DMC (pres.\ wk.)               & $3$&$.495$   & $11$&$.5(9)$ \\

AB  & DFT-LCAO-OO \cite{Dappe_2012} & $3$&$.1$--$3.2$ & ~~~~~~$70$&$(5)$ \\

AB  & SAPT(DFT) \cite{Podeszwa_2010} & $3$&$.43$   & $42$&$.5$    \\

AB  & vdW-DF \cite{Chakarova_2006}  & $3$&$.6$        & $45$&$.5$    \\

AB  & vdW-DF \cite{Lebedeva_2011}   & $3$&$.35$    & $29$&$.3$    \\

AB  & DFT-D \cite{Lebedeva_2011}    & $3$&$.25$    & $50$&$.6$    \\

AB  & DFT-D \cite{Gould_2013}       & $3$&$.32$    & $22$&     \\

AB  & DMC (pres.\ wk.)              & $3$&$.384$   & $17$&$.7(9)$ \\

\hline \hline
\end{tabular}
\end{table}

We have used the variational quantum Monte Carlo and DMC methods as
implemented in the \textsc{casino} code \cite{Needs_2010} to study MLG and
BLG\@. In the former method, Monte Carlo integration is used to evaluate
expectation values with respect to trial many-body wave-function forms that
may be of arbitrary complexity. In the DMC method
\cite{Ceperley_1980,Foulkes_2001}, a stochastic process governed by the
Schr\"{o}dinger equation in imaginary time is simulated to project out the
ground-state component of the trial wave function. Fermionic antisymmetry is
maintained by the fixed-node approximation, in which the nodal surface is
constrained to equal that of the trial wave function \cite{Anderson_1976}.
DMC methods have recently been used to study the BE of hexagonal boron nitride
bilayers \cite{Hsing_2014}.

Our many-body trial wave-function form consisted of Slater determinants for
spin-up and spin-down electrons multiplied by a symmetric, positive Jastrow
correlation factor $\exp(J)$ \cite{Foulkes_2001}. The Slater determinants
contained Kohn-Sham orbitals that were generated using the \textsc{castep}
plane-wave DFT code \cite{Clark_2005} within the local density approximation
(LDA)\@.  We performed test DMC calculations for $3 \times 3$ supercells of
MLG and AB-stacked BLG using Perdew-Burke-Ernzerhof (PBE) \cite{Perdew_1996}
orbitals.  The effect of changing the orbitals on the DMC total energies (and
hence the BE) was statistically insignificant.

To improve the scaling of our DMC calculations and to allow the use of
2D-periodic boundary conditions, the orbitals were re-represented in a
B-spline (blip) basis \cite{Alfe_2004}. The Jastrow exponent $J$ consisted of
polynomial and plane-wave expansions in the electron--ion and
electron--electron distances \cite{Drummond_2004}. The free parameters in the
Jastrow factor were optimized by unreweighted variance minimization
\cite{Drummond_2005,Umrigar_1988a}. The DMC energy was extrapolated linearly
to zero time step and we verified that finite-population errors in our results
are negligible \cite{footnote_supplementary}. The fixed-node error is of
uncertain magnitude, but it is always positive, and should largely cancel when
the BE is calculated. We used Dirac-Fock pseudopotentials to represent the C
atoms \cite{Trail_2005a,Trail_2005b} and fixed the in-plane lattice parameter
at the experimental value of $a=2.460$ {\AA}.

The principal source of uncertainty in our BE results is the need to use
finite simulation cells subject to periodic boundary conditions in DMC
calculations for condensed matter. Finite-size errors in DMC total energies
consist of (i) pseudorandom, oscillatory single-particle finite-size errors
due to momentum quantization and (ii) systematic finite-size errors due to the
inability to describe long-range two-body correlations and the difference
between $1/r$ and the 2D Ewald interaction \cite{Parry,Wood} in a finite
periodic cell.  By dividing the electron--electron interaction energy into a
Hartree term (the electrostatic energy of the charge density) and an
exchange--correlation energy (the interaction energy of each electron with its
accompanying exchange--correlation hole) and considering the long-range
nonoscillatory behavior of the hole predicted by the RPA, it can be shown that
the systematic finite-size error in the interaction energy per electron of a
2D-periodic system is negative and scales asymptotically with system size $N$
as $O(N^{-5/4})$ \cite{Drummond_2008}. The leading-order long-range
finite-size error in the kinetic energy per electron behaves in a similar
fashion.  The finite-size error in the atomization energy is therefore
positive and scales as $O(N^{-5/4})$, and the finite-size error in the BE per
atom must also exhibit the $O(N^{-5/4})$ scaling.  We also investigated
finite-size errors in the asymptotic BE using the Lifshitz theory of vdW
interactions \cite{Gomez_2009,Klimchitskaya_2013} with a Dirac model of
electron dispersion in graphene. To study finite system sizes, we introduced a
cutoff wavelength that depended on the cell size and layer separation.
However, near the equilibrium separation, short-range interactions are
important and the contribution to the finite-size error from the Lifshitz
theory is negligible. In order to eliminate finite-size effects and obtain the
atomization and binding energies in the thermodynamic limit, we studied
simulation cells consisting of arrays of $3 \times 3$, $4 \times 4$, and $6
\times 6$ primitive cells for MLG and BLG at the equilibrium layer separation
and $3 \times 3$ and $5 \times 5$ cells for BLG at nonequilibrium layer
separations. We used canonical-ensemble twist averaging \cite{lin_twist_av}
(i.e., averaging over offsets to the grid of ${\bf k}$ vectors) to reduce the
oscillatory single-particle finite-size errors in the ground-state energies of
MLG and BLG\@. To obtain the twist-averaged energy of MLG in a simulation cell
containing $N_P$ primitive cells, we performed DMC calculations at twelve
random offsets ${\bf k}_s$ to the grid of ${\bf k}$ vectors, then fitted
\begin{equation}
E(N_P,{\bf k}_s)=\bar{E}(N_P)+b[E_{\rm LDA}(N_P,{\bf k}_s)-E_{\rm
    LDA}(\infty)]
\label{eq:twist-averaged}
\end{equation}
to the DMC energies per atom $E(N_P,{\bf k}_s)$. The model function has two
fitting parameters: $\bar{E}(N_P)$, which is the twist-averaged DMC energy per
atom, and $b$. $E_{\rm LDA}(N_P,{\bf k}_s)$ is the DFT-LDA energy per atom of
MLG obtained using the offset ${\bf k}$-point grid corresponding to the
supercell used in the DMC calculations, and $E_{\rm LDA}(\infty)$ is the
DFT-LDA energy per atom obtained using a fine ($50 \times 50$) ${\bf k}$-point
mesh. Finally, we extrapolated our total-energy data to infinite system size
by fitting
\begin{equation}
\bar{E}(N_P)=E(\infty)+cN_P^{-5/4}
\label{eq:extrapolate}
\end{equation}
to the twist-averaged energies per atom, where the extrapolated energy per
atom $E(\infty)$ and $c$ are fitting parameters.  The atomization energy of
MLG is the difference between the energy of an isolated, spin-polarized C atom
and the energy per atom of MLG\@.

\begin{figure}
\begin{center}
\includegraphics[clip,scale=0.33]{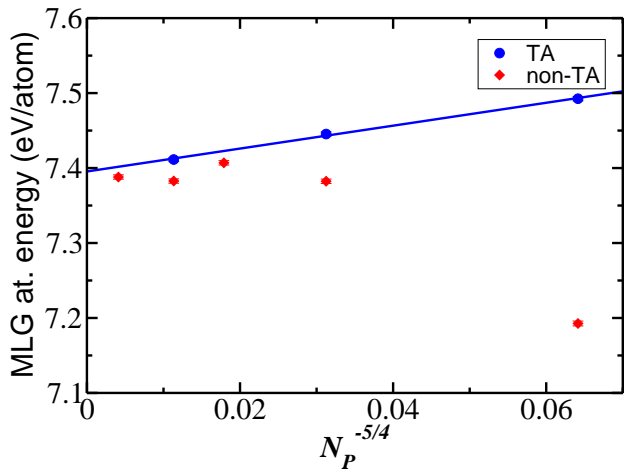}
\caption{(Color online) Twist-averaged (TA) and non-TA atomization energies of
  MLG against $N_P^{-5/4}$ as calculated by DMC, where $N_P$ is the number of
  primitive cells in the simulation supercell.
\label{fig:atomization_extrap}}
\end{center}
\end{figure}

Our DMC atomization energies of MLG as a function of system size are plotted
in Fig.\ \ref{fig:atomization_extrap}. We find the static-nucleus DMC
atomization energy to be $7.395(3)$ eV/atom with a Slater--Jastrow trial wave
function. This is lower than the DMC result of $7.464(10)$ eV/atom reported in
Ref.\ \cite{Shin_2014}. Most of this disagreement arises from the use of
different pseudopotentials in the two works \cite{footnote_supplementary}. The
DFT-PBE phonon zero-point energy (ZPE) of MLG was calculated using the method
of finite displacements in a $6\times6$ supercell \cite{Refson_2006} and found
to be $0.165$ eV/atom.  The ZPE is a correction to be subtracted from the
static-nucleus atomization energy. In principle, an accurate first-principles
atomization energy for graphene could be used to estimate the BE of graphite
by taking the difference of the experimental atomization energy of graphite
[$7.371(5)$ eV/atom \cite{CRC_handbook}] and the ZPE-corrected atomization
energy of MLG\@. However, the spread of DFT atomization energies resulting
from different choices of pseudopotential (of order 40--70 meV/atom
\cite{footnote_supplementary}) implies that first-principles pseudopotential
calculations cannot currently be used to calculate the BE of graphite by this
approach.

Despite a great deal of theoretical and experimental work, the BE of graphene
layers remains poorly understood. The cleavage energy of graphite has been
measured to be $43(5)$ meV/atom \cite{Girifalco_1956}, the BE to be $35(10)$
meV/atom \cite{Benedict_1998}, and the exfoliation energy to be $52(5)$
meV/atom \cite{Zacharia_2004}. More recent experimental work has found the
cleavage energy to be $31(2)$ meV/atom \cite{Liu_2012}. It has been suggested
that the latter result may be substantially underestimated, because the
experimental data were analyzed using a Lennard-Jones potential, which gives
qualitatively incorrect interlayer BEs at large separation
\cite{Gould_2014}. Similar difficulties of interpretation may affect the other
experimental results in the literature.  The results obtained in these works
are widely scattered.  The DMC method has previously been applied to calculate
the BE of graphite \cite{Spanu_2009}, which was found to be 56(5) meV/atom,
although these calculations were performed in relatively small simulation
supercells, and finite-size effects may limit the accuracy of the results
obtained.

\begin{figure}
\begin{center}
\includegraphics[clip,scale=0.33]{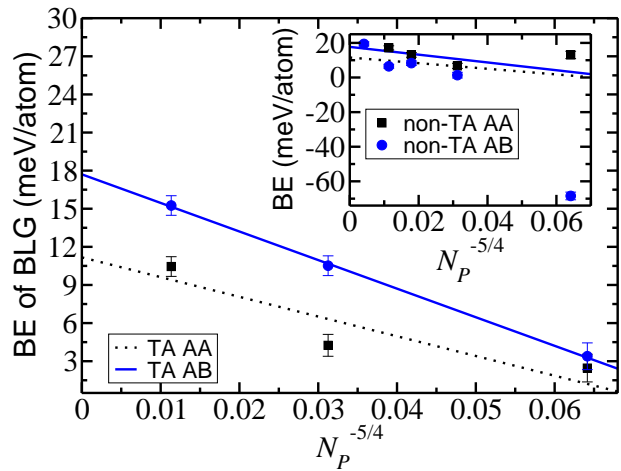}
\caption{(Color online) Twist-averaged (TA) BLG BE against $N_P^{-5/4}$ as
  calculated by DMC, where $N_P$ is the number of primitive cells in the
  simulation supercell. The inset shows non-twist-averaged BEs.  The layer
  separations are the vdW-DF \cite{Brihuega_2012} equilibrium values of
  $3.495$ and $3.384$ {\AA} for the AA- and AB-stacked structures,
  respectively.
 \label{fig:binding_extrap_equil}}
\end{center}
\end{figure}

For BLG, we restrict our attention to the nonretarded regime
\cite{footnote_nonretarded}, in which the BE is simply the difference between
the nonrelativistic total energy per atom in the monolayer and the bilayer.
We used vdW-DF layer separations of $d=3.495$ {\AA} and $3.384$ {\AA}
\cite{Brihuega_2012} for the AA- and AB-stacked configurations,
respectively. In Fig.\ \ref{fig:binding_extrap_equil} we plot the
twist-averaged BEs of AA- and AB-stacked BLG as a function of system
size. Non-twist-averaged BEs are shown in the inset to
Fig.\ \ref{fig:binding_extrap_equil} and, as expected, show large oscillations
due to momentum-quantization effects. For widely separated graphene layers
with nonoverlapping charge densities, single-particle finite-size errors
cancel perfectly when the BE is calculated. However, when the layers are
closer together, the cancellation is no longer perfect. In practice, near the
equilibrium separation, the single-particle errors in the BE correlate closely
with the single-particle errors in the total energy of BLG\@.  To evaluate the
BE in the thermodynamic limit, we twist-averaged the BE using
Eq.\ (\ref{eq:twist-averaged}) with the BE per atom in place of $E(N_P,{\bf
  k}_s)$ and the DFT-LDA total energy per atom of BLG in place of $E_{\rm
  LDA}(N_P,{\bf k}_s)$.  We then extrapolated the twist-averaged BE to
infinite system size using Eq.\ (\ref{eq:extrapolate}). As shown in
Fig.\ \ref{fig:binding_extrap_equil}, the BE of AB-stacked BLG is larger than
that of AA-stacked BLG, confirming that the former is the more stable
structure.

The area of a simulation cell with $N_P$ unit cells is $A=\sqrt{3}N_Pa^2/2$,
where $a$ is the lattice parameter of graphene.  If we define the linear size
$L$ of the cell via $\pi L^2=A$ then we may express the twist-averaged BE per
atom as $\bar{E}_{\rm bind}(L)=E_{\rm bind}(\infty)+c'L^{-5/2}$, where $c'$ is
$-0.31(5)$ and $-0.43(5)$ eV\,{\AA}$^{5/2}$ for the AA-stacked and AB-stacked
geometries, respectively.  The BE is reduced at small supercell sizes $L$.
The use of a finite supercell crudely models the situation where the Coulomb
interaction between electrons is screened by a metallic substrate.  Hence a
metallic substrate is expected to weaken the binding of BLG\@.

\begin{figure}
\begin{center}
\includegraphics[clip,scale=0.33]{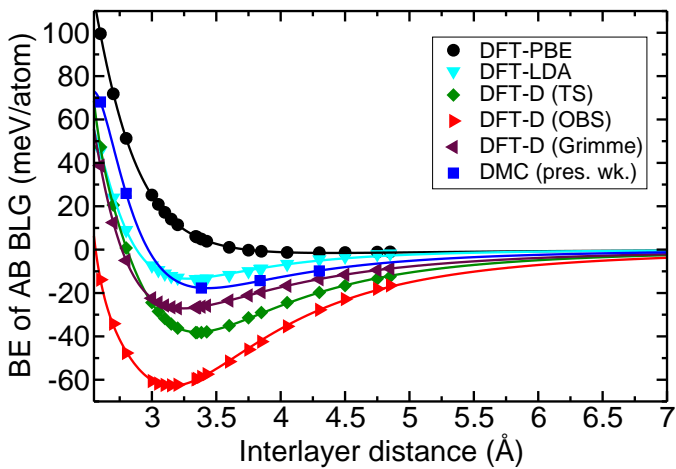}
\caption{(Color online) BE curve of AB-stacked BLG as a function of interlayer
  distance calculated using DFT, DFT-D, and DMC methods.  Our DFT-D
  calculations used the Tkatchenko--Scheffler (TS) \cite{Tkatchenko_2009},
  Ortmann--Bechstedt--Schmidt (OBS) \cite{OBS_2006}, and
  Grimme~\cite{Grimme_2006} vdW corrections.
\label{fig:binding_curve}}
\end{center}
\end{figure}

In Fig.\ \ref{fig:binding_curve} we plot the BE of AB-stacked BLG against the
interlayer separation, as calculated by DFT, DFT-D, and DMC\@. The layer
separations we have studied are not in the asymptotic regime in which the BE
falls off as $d^{-3}$, where $d$ is the interlayer separation
\cite{Dobson_2014}. We have fitted the function $E_{\rm
  bind}(d)=A_4d^{-4}+A_8d^{-8}+A_{12}d^{-12}+A_{16}d^{-16}$ to our DMC BE
data, where the $\{A_i\}$ are fitting parameters, which we find to be
$A_4=-2.9\times 10^3$ meV\,{\AA}$^4$, $A_8=-2.97\times 10^5$ meV\,{\AA}$^8$,
$A_{12}=6.18 \times 10^7$ meV\,{\AA}$^{12}$, and $A_{16}=-1.63\times 10^9$
meV\,{\AA}$^{16}$.  This function fits the DMC data well, with a $\chi^2$
value of $0.007$ per data point. The BE found at the minimum of the fitting
curve is $17.8(8)$ meV/atom at the equilibrium separation of $3.43(4)$ \AA\@.
Although the separation that minimizes our fitted BE curve for AB-stacked BLG
is somewhat larger than the separation used in our calculation of the BE
reported in Table \ref{table:compare_BE}, the difference between the BEs is
not statistically significant.  The Tkatchenko--Scheffler
\cite{Tkatchenko_2009} DFT-D scheme shows roughly the same equilibrium
separation as DMC, but the magnitude of the BE is substantially larger. In
general, the three DFT-D methods studied
\cite{Tkatchenko_2009,OBS_2006,Grimme_2006} disagree with each other and with
DMC\@.  Indeed, the magnitude of the BE (if not the shape of the BE curve) is
best described by the LDA\@.  Our fitted BE curve enables us to calculate the
out-of plane zone-center optical phonon frequency $\omega_{{\rm ZO}'}$ of
AB-stacked BLG \cite{footnote_freq}. A comparison of $\omega_{{\rm ZO}'}$
frequencies obtained by DFT, DMC, and experiments \cite{Lui_2013,Milana_2015}
is shown in Table \ref{table:phonon_freq}.  Our DFT-LDA frequency is in
reasonable agreement with the result (76.8 cm$^{-1}$) reported in
Ref.\ \cite{Saha_2008}.  The difference between the $\omega_{{\rm ZO}'}$
frequency predicted by our fit to our DMC data and the experimental result is
negligible [$3(7)$ cm$^{-1}$] \cite{footnote_supplementary}.

\begin{table}
\caption{The equilibrium separation $d_0$, static-lattice BE at
  equilibrium separation, and out-of-plane zone-center optical-phonon
  frequency $\omega_{{\rm ZO}'}$ of AB-stacked BLG\@ obtained by DFT, DFT-D,
  DMC, and experiment.  The minimum of the curve fitted to the DMC BE data,
  which is reported in this table, is in statistical agreement with the DMC BE
  obtained using a fixed layer separation of 3.384 {\AA}, which is reported in
  Table \ref{table:compare_BE}.}
\label{table:phonon_freq}
\begin{tabular}{lr@{}lr@{}ll} \hline \hline
    Method       &  \multicolumn{2}{c}{$d_0$ (\AA)}   &
     \multicolumn{2}{c}{BE (meV/at.)}& $\omega_{{\rm ZO}'}$
       (cm$^{-1}$)  \\ \hline

        DFT-PBE  & $4.$&$40$ & $1.$&$53$  & $16$ \\
        DFT-LDA  & $3.$&$28$ & $13.$&$38$ &$84$  \\
        DFT-D (TS)& $3.$&$35$ & $38.$&$03$ &$111$  \\
        DFT-D (OBS)& $3.$&$15$ & $62.$&$70$ &$133$ \\
        DFT-D (Grimme)& $3.$&$25$ & $27.$&$08$ & $95$ \\
        DMC (pres.\ wk.) & $3.$&$43(4)$& $17.$&$8(8)$ &$83(7)$ \\
        Exp.~\cite{Lui_2013} & &&      &       & $80 (2)$  \\ 
        Exp.~\cite{Milana_2015} & &&      &       & $89 $  \\ \hline \hline
 \end{tabular}
\end{table}

In summary, we have used the DMC method to determine the BE of BLG\@. Our
approach includes a full, first-principles treatment of vdW interactions. We
have found the static-nucleus atomization energy of MLG to be $7.395 (3)$
eV/atom, although the uncertainty in this result due to the use of nonlocal
pseudopotentials may be as much as 70 meV/atom
\cite{footnote_supplementary}. We find the BEs of AA- and AB-stacked BLG near
their equilibrium separations to be $11.5(9)$ and $17.7(9)$ meV/atom,
respectively. Our results indicate that current DFT-D and vdW-DF methods
significantly overbind 2D materials.

\begin{acknowledgments}
We acknowledge financial support from the UK Engineering and Physical Sciences
Research Council (EPSRC)\@. This work made use of the facilities of Lancaster
University's High-End Computing facility, N8 HPC provided and funded by the N8
consortium and EPSRC (Grant No.\ EP/K000225/1), and the ARCHER UK National
Supercomputing Service. We acknowledge discussions with Andrea C. Ferrari and
Silvia Milana.
\end{acknowledgments}


\begin{thebibliography}{99}
\bibitem{footnote_freq} The interlayer BE per atom of BLG can be
  written as $E_{\rm bind}(d)=E_0+\frac{1}{8}m_C\omega_{{\rm ZO}'}^2
  (d-d_0)^2+O(d-d_0)^3$, where $E_0$ is the BE per atom at the
  equilibrium separation $d_0$, $m_C$ is the mass of a carbon atom, and
  $\omega_{{\rm ZO}'}$ is the out-of-plane phonon frequency.
\end{thebibliography}
\end{document}